\documentclass[prd,nofootinbib,twocolumn,
showpacs,preprintnumbers]{revtex4}
\usepackage{graphicx}
\usepackage{slashed}
\usepackage{amssymb}
\usepackage{bm}

\newcommand{\intsum}{{\scriptstyle\sum}\hspace{-11pt}\int}
\newcommand{\textintsum}{{\scriptstyle\Sigma}\hspace{-8pt}\int}
\newcommand{\ttin}[1]{\textrm{\tiny #1}}

\begin{document}

\title{Weinberg power counting and the quark determinant at small chemical potential}

\author{E. S. Fraga$^{1}$ and C. Villavicencio$^{2}$}
\affiliation{
$^{1}$Instituto de F\'{\i}sica, Universidade Federal do Rio de Janeiro, 
C.P. 68528, Rio de Janeiro, RJ 
21945-972, Brasil  \\
$^{2}$Facultad de F\'\i sica, Pontifcia Universidad Cat\'olica de Chile, Casilla
306, Santiago 22, Chile}

\begin{abstract}
We construct an effective action for QCD by expanding the quark 
determinant in powers of the chemical potential at finite temperature in 
the case of massless quarks. To cut the infinite series we adopt the Weinberg 
power counting criteria. We compute the minimal effective action ($\sim p^4$), 
expanding in the external momentum, which implies the use of the hard thermal 
loop approximation. Our main result is a gauge invariant expression for the 
phase $\theta$ of the functional determinant in QCD, and recovers dimensional reduction 
in the high-temperature limit. We compute, analytically,
$\langle\theta^{2}\rangle$ in 
the range of $p\ll 2\pi T$, including perturbative and nonperturbative
contributions, 
the latter treated within the mean field approximation. Implications for lattice simulations 
are briefly discussed.
\end{abstract}


\maketitle
\section{Introduction}

There has been an increasing interest in the last few years in the
\emph{sign problem} or \emph{phase problem} in QCD \cite{lattice2009}. 
For a finite chemical potential, $\mu$, the fermion determinant matrix is
non-positive definite, so it is not possible to perform Monte Carlo 
simulations in the usual fashion \cite{Laermann:2003cv,Hands:2007by}.
Nevertheless, the Glasgow method \cite{Barbour:1991vs} and rewheighting 
techniques \cite{Fodor:2001au} have provided great advances in the description 
of phase transitions on the lattice, considering  a set of parameters near 
the transition line. 

There is special interest in the region of high temperature and low chemical 
potential, since it corresponds to the sector of the phase diagram of strong 
interactions probed by high-energy heavy-ion collision experiments
\cite{QM2009}. 
In this regime it is possible to expand the fermion determinant in powers 
of $\mu/\Lambda$, where $\Lambda$ is some mass scale related to the temperature, 
$\Lambda\sim T$ \cite{Gottlieb:1987ac,Ejiri:2002up,Allton:2003vx,Gavai:2004sd,Allton:2005gk}. 
In association with the ones mentioned previously, this technique is very convenient 
and successful to describe this region. Besides, several other complementary approaches 
were proposed with the intent to shed some light on the sign problem 
\cite{de Forcrand:2002ci,Hong:2003zq,Imachi:2006mw,Fukushima:2006uv,Splittorff:2007ck,Han:2008xj}.

In this paper we investigate the small chemical potential 
sector of the QCD phase diagram. In particular, we present a scheme that is 
valid for a wide range of temperatures in the soft region. Methods that are 
based on an expansion in $\mu$ always have to resort to approximations to 
compute the coefficients of the series. For this purpose, there are numerical 
approximation techniques that yield good results but do not allow for a deeper 
analytic study, the most frequently used being the random noise 
method (see, e.g., Ref. \cite{Allton:2002zi}). Analytic treatments are usually 
restricted to very high temperatures. For instance, the dimensional reduction 
effective action \cite{Ginsparg:1980ef,Appelquist:1981vg,Braaten:1995jr,
Kajantie:1995dw,Hart:2000ha,Vuorinen:2003fs} is valid for $T \gtrsim 2T_c$, 
where $T_c$ is the critical temperature. A complete study of the pressure
within perturbative QCD for all values of the temperature and chemical potential 
can be found in Ref. \cite{Ipp:2006ij} where the authors make use of the hard thermal 
loop, hard dense loop and dimensional reduction approximations. 

Here we start from a different perspective: instead of expanding all quantities 
and desired observables in powers of $\mu/\Lambda$, the idea is to keep the 
relevant terms in the effective action according to the Weinberg power counting 
criteria \cite{Weinberg:1978kz}. In what follows we construct an effective
action 
for QCD by expanding the quark determinant in powers of the chemical potential 
at finite temperature in the case of massless quarks. We compute the minimal 
effective action expanding in the external momentum up to order $\sim p^4$ in 
power counting. In practice, the momentum expansion performed here is equivalent 
to the hard thermal loop (HTL) approximation 
\cite{Braaten:1989mz,Braaten:1990az,Frenkel:1989br}. 
Our main result is a gauge invariant expression for the phase angle $\theta(\mu)$ of the 
functional determinant in QCD, which can be written as 
$\det M(\mu)=|\det M(\mu)|~e^{i\theta(\mu)}$. 
An interesting analysis of the angle 
$\theta$ has been recently performed using the random matrix framework 
\cite{Splittorff:2007ck,Han:2008xj}. 

As a first test of our method, we recover the dimensional reduction approximation 
in the limit of high temperature. As a second step, we compute analytically 
$\langle\theta^{2}\rangle$ in the range of $p\ll 2\pi T$, keeping nonzero mode 
contributions in the effective action. In this calculation, we include perturbative 
and nonperturbative contributions, treating the latter in the mean field
approximation. 
Implications for lattice simulations are also discussed. 

This paper is organized as follows. In Sec. II. we develop the general
expansion for 
the fermionic determinant,  present the Feynman rules and the relevant diagrams, 
settling the framework. In Sec. III. we discuss the power counting hierarchy in
order 
to be able to cut the series and compute the minimal effective action. As a first test of 
our framework, we also reobtain the result from dimensional reduction in the limit of high 
temperature, including the non-zero mode terms. In Section IV we present the calculation 
of $\langle\theta^{2}\rangle$ in the range of $p\ll 2\pi T$. Section V contains our conclusions.

\section{The expansion}

\noindent
The generating functional for QCD with massless quarks at finite chemical potential 
is defined, in euclidean space, as
\begin{equation}
{\cal Z}=\int {\cal D}G
\det(-i\slashed{D}+i\mu\gamma_4)e^{-S_{\mbox{\tiny YM}}[G]} \; ,
\end{equation}
where $G$ are the gluon fields, also present in the covariant
derivative $D_{\mu}=\partial_{\mu}-iG_{\mu}$,
and $S_{\mbox{\tiny YM}}$ is the Yang-Mills (YM) action. 
We can expand the fermion determinant in powers of the chemical
potential
assuming that $\mu<\Lambda\sim T$:
\begin{eqnarray}
&&\det (-i\slashed{D}+i\mu\gamma_4)=\nonumber\\
&&~~~
\det(-i\slashed{D})\exp\Bigg\{-N_f\sum_{s=1}^\infty\frac{(-i\mu)^s}{s
} 
\int_\beta dy_{1}\cdots dy_{s}
\nonumber\\
&&~~~~
\mbox{Tr}~\gamma_4S(y_{2},y_{1})\gamma_4S(y_{3},y_{2})\cdots
\gamma_4S(y_{1},y_{n})\Bigg\},
\label{expan_mu}
\end{eqnarray}
where $\int_\beta dy\equiv\int_0^\beta dy_4\int d^3y$,
$S(y_{b},y_{a})$ is the dressed fermion propagator, which
can be
expressed as
a series in powers of the gauge field and the free fermion
propagator using the self-consistent relation
\begin{equation}
S(x,y)=S_F(x-y)-\int_\beta dz S(x,z)\slashed{G}(z)S_F(z-y) \; .
\end{equation}

The expansion, then, will contribute to additional terms in the
effective action $S_{\mbox{\tiny eff}}
=S_{\mbox{\tiny YM}}
+\sum_{n,s} S^{(n,s)}$. The new terms, expressed in 
momentum space, are of the form
\begin{eqnarray}
S^{(n,s)} &=&
\mu^s\intsum dp_{1}\cdots dp_{n}
  (2\pi)^4\delta(p_1+\cdots +p_n)
\nonumber\\
&&
\Gamma^{(s,n)} _{\mu_1\cdots\mu_n}(\{p_{i}\})
~\textrm{tr}\tilde G_{\mu_1}(p_{1})\cdots \tilde
G_{\mu_n}(p_{n}) \, ,
\label{Seff_YM}
\end{eqnarray}
where the last integral denotes also the sum over bosonic Matsubara
frequencies $\textintsum dp\equiv
T\int\frac{d^3p}{(2\pi)^3}\sum_{p_4=2n_p\pi T}$, and $2\pi
T\delta(p_4)=\delta_{n_p,0}$. 
In this way, one obtains a positive-definite fermion determinant, and
the
contribution from the chemical potential will be part of an effective
gluon action.

\begin{figure}
\includegraphics[scale=.4]{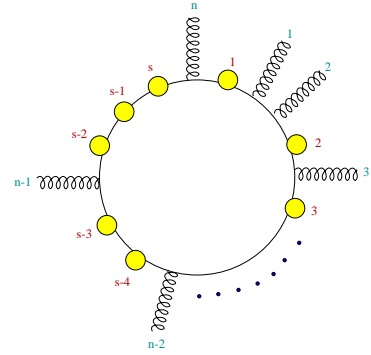} 
\caption{General diagram for the construction of the effective vertices.
The
small circles correspond to chemical potential insertions.}
\label{Gns1}
\end{figure}

The Feynman rules in momentum space for calculating the different
effective vertices are almost the same as the usual ones. 
The 
difference is that all of the operators between chemical potential 
insertions must be transposed in order. Figure \ref{Gns1} shows 
a general diagram with chemical potential and gluon insertions.
To construct a diagram for a vertex with $n$ gluons and $s$
chemical potential insertions, one puts 
$-t_{a_i}\gamma_{\mu_i}$ for any gluon insertion, $-i\gamma_4$
for any chemical potential insertion, and divides by the symmetry 
factor $s$.

\begin{figure}
\hspace{1.5cm}\includegraphics[scale=.49]{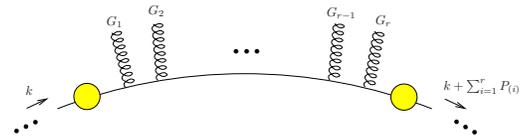} 

\caption{A set of operators between two chemical potential insertions.}
\label{Gns2}
\end{figure}

Between chemical
potential insertions, the order of the operator must be transposed considering
momentum conservation.
In a piece of the effective vertex shown in
Fig. \ref{Gns2}, the integrand must be written as
\begin{eqnarray}
&&
(-i\gamma_4)
\bigg[\tilde S_F(k)(-t_{a_1}\gamma_{\mu_1})\tilde
S_F(k+p_{1})
\cdots
\nonumber\\&&\qquad
\cdots (-t_{a_r}\gamma_{\mu_r})\tilde
S_F(k+p_{1}+\cdots+p_{r})\bigg]^{(t)}
(-i\gamma_4) \; ,\qquad 
\end{eqnarray}
where the exponent $(t)$ in the brackets is a reminder to 
transpose the order of  the operators:
$
[{\cal O}_1{\cal O}_2\cdots {\cal O}_{r-1}{\cal O}_r]^{(t)}
={\cal O}_{r}{\cal O}_{r-1}\cdots {\cal O}_2{\cal O}_1
$.

Finally, one takes the trace over gamma matrices and color group
representation, 
integrating over internal fermionic momentum (odd Matsubara
frequencies).
The sum of all diagrams will produce the effective vertices
$\Gamma^{(n,s)}$, 
which will be invariant under any cyclic change in the set of indices
$\mu_i$,
$a_i$, $p_{i}$ (or $x_{i}$ in the
case of configuration space).

%

\section{The minimal effective action}

Now, we need a criteria to cut the series. For low-energy effective theories, 
one can consider the Weinberg power counting \cite{Weinberg:1978kz}, which 
uses the argument that all mass parameters (external momentum, chemical potential 
and gluon fields) must be less than a certain scale that is proportional to the
temperature. This approach has provided a very successful description within chiral
effective models of QCD. 
Although this is not exactly the case here, it is reasonable 
that for low-energy processes at high temperature typical values of the
operators 
mentioned above, as well as the chemical potential, can be considered to be in a
region 
of the same order or smaller than the scale. Then, assuming $\mu\sim G\sim p$, and 
expanding the effective Lagrangean in soft modes, we can cut it at a given order 
in powers of the momentum scale.

In the case of Yang-Mills theories, this soft-mode expansion for high temperatures 
corresponds to the HTL approximation. The minimal action must be of
order $(p)^4$. 
So, applying the power counting criteria, the minimal effective action is given
by 
$ S_\ttin{eff}^\ttin{min}= S_\ttin{YM}
+S^{(0,2)}+S^{(0,4)}+ \big[S^{(2,2)}+S^{(3,1)}\big]_{{\cal O} (p)^0}$,
where the indices $(n,s)$ are defined in Eq. (\ref{Seff_YM}), and the last two
terms are expanded in momentum up to zeroth order which leads to the appearance 
of functions of $|\bm{p}|/p_4$.

The whole series of gauge fields is gauge invariant at each order in the expansion 
in $\mu$, i.e. $\sum_nS^{(n,s)}$ is gauge invariant for all values of
$s$ as can be 
seen directly from Eq. (\ref{expan_mu}). Since the terms in the sum are traces 
containing dressed propagators, which are gauge invariant, every term in the sum 
is gauge invariant. 
Moreover, the minimal effective vertices that we need satisfy Ward 
identities of the form 
\begin{eqnarray}
p_\mu \Gamma^{(2,2)}_{\mu\nu}(p)&=&0 \; ,\\
p_\mu \Gamma^{(3,1)}_{\mu\nu\sigma}(p,q,r)&=&
\Gamma^{(2,1)}_{\nu\sigma}(r)-
\Gamma^{(2,1)}_{\nu\sigma}(q) \; , 
\label{ward}
\end{eqnarray}
and analogous relations obtained by changing cyclical indices and arguments. 
Equation (\ref{ward}) vanishes, since $\Gamma^{(2,1)}=0$. As is well
known, 
HTL preserves the Ward identities.


The nonvanishing diagrams for the vacuum contributions
$\Gamma^{(s,0)}$ are the known vacuum corrections to the
thermodynamic potential
\begin{equation}
\Gamma^{(0,2)}=-N_cN_f\frac{T^2}{6}
\; , \qquad
\Gamma^{(0,4)}=-N_cN_f\frac{1}{12\pi^2} \; .
\label{G0s}
\end{equation}
For $s>4$ all contributions vanish, as was demonstrated in 
Ref. \cite{KorthalsAltes:1999cp}. The next nonvanishing term has 
the form of the polarization tensor in the HTL approximation
\begin{equation}
\Gamma^{(2,2)}_{\mu\nu}(p)=\frac{N_f}{2\pi^2}\int\frac{d\Omega}{
4\pi }
\left[
\frac{ip_4}{\hat k\cdot p}\hat k_\mu\hat k_\nu+\delta_{\mu 4}\delta_{\nu 4}
\right] \; ,
\end{equation}
with the lightlike four-vector $\hat k=(\hat{\bm{k}},i)$. Finally, the
vertex components which correspond to $i\theta$ are
\begin{widetext}
\begin{eqnarray}
\Gamma_{\mu\nu\rho}^{(3,1)}(p,q,r) =
\frac{i N_f}{6\pi^2}\int\frac{d\Omega}{4\pi}\Bigg\{
 2\delta_{\mu 4}\delta_{\nu 4}\delta_{\rho 4}
-\frac{p_4}{\hat k\cdot p}\Big[
  \hat k_\mu\delta_{\nu\rho}
  -6\hat k_\mu \hat k_\nu \hat k_\rho  
  +2i\left(\hat k_\mu\hat k_\nu \delta_{\rho 4}
    +\hat k_\nu\hat k_\rho \delta_{\mu 4}
    +\hat k_\rho\hat k_\mu \delta_{\nu 4}\right)\Big]
\nonumber\\
-\frac{q_4}{\hat k\cdot q}\hat k_\nu\delta_{\rho\mu}    
-\frac{r_4}{\hat k\cdot r}\hat k_\rho\delta_{\mu\nu}   
+2i\left(\frac{p_4}{\hat k\cdot p}\right)^2
  \hat k_\mu \hat k_\nu \hat k_\sigma
\nonumber\\
+\frac{q_4}{\hat k\cdot q~\hat k\cdot p}\Big[
  2iq_4\hat k_\mu\hat k_\nu\hat k_\rho
  +\left(q_\nu\hat k_\rho-r_\rho \hat k_\nu\right)\hat k_\mu
  +\left(q_\mu-r_\mu\right)\hat k_\nu\hat k_\rho\Big]
\nonumber\\
+\frac{r_4}{\hat k\cdot r~\hat k\cdot p}\Big[
  2ir_4\hat k_\mu\hat k_\nu\hat k_\rho  
  +\left(r_\rho\hat k_\nu-q_\nu\hat k_\rho\right)\hat k_\mu
  +\left(r_\mu-q_\mu\right)\hat k_\nu\hat k_\rho\Big]
\nonumber\\
+\Bigg[\frac{q_4(q^2-r^2)}{\hat k\cdot q~(\hat k\cdot p)^2}
  -\frac{q_4q^2}{(\hat k\cdot q)^2~\hat k\cdot p}
  +\frac{r_4(r^2-q^2)}{\hat k\cdot r~(\hat k\cdot p)^2}
  -\frac{r_4r^2}{(\hat k\cdot r)^2~\hat k\cdot p}\Bigg]
    \hat k_\mu\hat k_\nu\hat k_\rho\Bigg\} \; .
\label{Gijm}
\end{eqnarray}
\end{widetext}


The dimensional reduction approximation can be obtained directly from the last expressions 
of the effective vertices by simply considering the case in which
$|\bm{p}|\ll 2\pi T$. In order to 
expand in powers of the external momentum we have to separate the zero mode from the other 
modes in the gluon fields $G_\mu=A_\mu+B_\mu$, with
\begin{eqnarray}
A_\mu(\bm{x}) &=& T\int\frac{d^3p}{(2\pi)^3} 
~e^{i\bm{p}\cdot\bm{x}} ~\tilde G_\mu(\bm{p},0) ,\\
B_\mu(x) &=& T\sum_{p_4\neq 0}\int\frac{d^3p}{(2\pi)^3} 
~e^{ip\cdot x} ~\tilde G_\mu(p),
\end{eqnarray}
$A_{\mu}$ being the field corresponding to the zero mode. The lowest order contribution 
in the expansion ($p\to 0$) is given by
\begin{eqnarray}
S^{(2,2)} &=& \frac{\mu^2 N_f}{2\pi^2}
\int_\beta dx ~\textrm{tr}\left[
A_4^2-\frac{1}{3}\bm{B}^2\right] ,
\label{G22dimred} \\
S^{(3,1)} &=& \frac{i\mu N_f}{3\pi^2}
\int_\beta dx ~\textrm{tr}\left[
A_4^3+A_4B_4^2\right]
\label{G31dimred}
\end{eqnarray}

If we set $B=0$, we recover the dimensional reduction effective action at tree 
level \cite{Hart:2000ha,Vuorinen:2003fs}. The usual effective action for the
high-temperature 
regime is constructed through loops integrating the $B$ fields. However, this is done perturbatively, 
and this is not the regime in  which we are interested. The fact of including $B$ as a nonperturbative 
field (except for the power counting) in principle will enhance the range of validity in temperature 
to values lower than in the case of just considering an infinite temperature expansion. 
In this sense, our framework goes beyond dimensional reduction, and can probe temperatures 
closer to $T_{c}$.

\section{$\langle\theta^{2}\rangle$ at high temperature}

The phase angle $\theta$ of the complex functional determinant is a crucial quantity for lattice 
simulations in QCD. The knowledge of $\theta$ allows for the separation of the functional 
integral into two different regions: $|\theta| \lessgtr \pi/2$. 
Nevertheless this is not a simple
task, since it is not restricted to values of $-\pi<\theta <\pi$ but seems to 
increase with the volume
\cite{Ejiri:2007ga}. The calculation of the average phase
\begin{equation}
\langle e^{2i\theta}\rangle =\left< \frac{\textrm{det}(-i\slashed{D}+i\mu\gamma_4)}
{\textrm{det}(-i\slashed{D}+i\mu\gamma_4)^\dag}\right>
\label{phaseaverage}
\end{equation}
gives a measure of how problematic could be the phase in lattice
simulations. As this quantity must vanish in the thermodynamic 
limit, it should be evaluated in a finite volume (otherwise the average phase
above vanishes as soon as $\mu$ is nonzero). 
The random matrix aproximation \cite{Splittorff:2007ck} yields a
vanishing result for 
$\mu>m_\pi/2$. One can use our effective expression for the phase to compute the average 
phase factor in Eq. (\ref{phaseaverage}). If one expands the 
exponential in powers of the angle, the relevant term will be $\sim \langle \theta^2\rangle$, 
since the average of the phase must be real.
In particular, the determination of $\langle \theta^2\rangle_{P}$, the average
of the angle as a function of the plaquette, is an important ingredient in order 
to localize the critical line in the temperature-number density phase diagram of 
QCD \cite{Ejiri:2007ga}.

In our framework, we have $S^{(3,1)}=i\theta$. Writing the fields in the form
\begin{eqnarray}
A_\mu^a(\bm{x})&=&T\int_0^\beta dx_4 G_\mu^a(x) \;, \\
B_\mu^a(x)&=&G_\mu^a(x)-T\int_0^\beta dx_4 G_\mu^a(x)\;,
\end{eqnarray}
we can use Eq. (\ref{G31dimred}) to express $\theta$ in the high-temperature limit as
\begin{eqnarray}
 \theta &=&
\frac{\mu N_f}{12\pi^2}d^{a b c}T
\int_V d^3x \int_0^\beta dx_4 dy_4
G_{\alpha}^{a}(y)
G_{\beta}^{b}(x)
G_{\gamma}^{c}(x) \;,\nonumber \\
\end{eqnarray}
where $x=(\bm{x},x_4)$, $y=(\bm{x},y_4)$ and $d^{a b c}$ are real and totally symmetric as usual. 
The expectation value of $\theta$ vanishes, so we compute the next power which can be expressed 
in terms of two-point correlation functions as
\begin{eqnarray}
 \hspace{-1cm}\langle \theta^2\rangle
&\approx&  
\frac{\mu^2 T^2 N_f^2(N_c^2-4)}{(12\pi^2)^2N_c(N_c^2-1)^2} 
\int_0^\beta dx_4 dx'_4 dy_4 dy'_4 
\nonumber \\&& \hspace{-1.1cm}\times
\int_V d^3x d^3x'
[2 \langle G_4^a(y)G_4^a(y')\rangle
\langle G_4^a(x)G_4^a(x')\rangle^2 \nonumber \\
&&   \hspace{-1.1cm} + 4 \langle G_4^a(y)G_4^a(x')\rangle
\langle G_4^a(x)G_4^a(y')\rangle
\langle G_4^a(x)G_4^a(x')\rangle]  
\end{eqnarray}  
where the expression is not exact since we allow for nonperturbative
contributions, 
and we used $\langle G^aG^b\rangle \sim\delta^{ab}$.

If we separate the gluon field in a perturbative and a nonperturbative
contribution, 
$G=G_{p}+G_{np}$, with $\langle G_pG_{np}\rangle=0$, there will be contributions
to 
$\langle \theta^2\rangle$ coming from a purely perturbative term, a purely
nonperturbative 
term and the crossed terms. The basic building blocks are the two-point
functions for 
the field strengths, which we compute in the sequel.

For the perturbative case, we start with the HTL effective Lagrangian
\begin{eqnarray}
{\cal L}_g&=&\frac{1}{4g^2}(G^a_{\mu\nu})^2 
+\frac{1}{4g^2}(\partial_\mu G_\mu^a)^2 \nonumber \\
&&+\frac{m^2}{4g^2}G^{a}_{\mu\alpha}\int\frac{d\Omega}{4\pi}\frac{\hat k_\alpha
\hat
k_\beta}{(i\hat k\cdot D)^2}G_{\beta\mu}^a \;,
\end{eqnarray}
where $m$ is the standard HTL effective mass \cite{lebellac}
\begin{equation}
m^2=\frac{g^2}{6}
\left[
T^2(N_f+2N_c)+\frac{3N_f\mu^2}{\pi^2}
\right] \;. 
\end{equation}
Then, the fourth component of the gluon two-point function can be written, in
the limit 
$p \ll 2\pi T$, as 
\begin{eqnarray}
\hspace{-1cm}\langle G_{4}^a(x)_pG_{4}^b(x')_p\rangle
&=&\delta^{ab}g^2T \int \frac{d^3p}{(2\pi)^3} 
e^{i\bm{p}\cdot(\bm{x}-\bm{x'})}\nonumber \\
&&\hspace{-1.1cm}\times \left[
\frac{1}{\bm{p}^2+m^2}
+ \sum_{n\neq 0} \frac{e^{i\omega_n (x_4 - x_4')}}{\omega_{n}^{2}+\bm{p}^2}
\right], 
\end{eqnarray}
where $\omega_{n}=2n\pi T$ is the bosonic Matsubara frequency.

For the nonperturbative case it is convenient to use the Schwinger gauge, also 
known as fixed point or coordinate gauge \cite{Pascual:1984zb}, 
$x\cdot G=0$. Then, the gauge field can be expressed in terms of the field
strength
tensor:
\begin{equation}
G^a_{\mu}(x)=\int_0^1 ds s  G^a_{\mu\alpha}(sx)x_\alpha \;.
\end{equation}
If we approximate the two-point function for the field strength by its mean
field 
value, i.e.
\begin{equation}
\langle G^a_{4\alpha}(x)G^a_{4\alpha}(y)\rangle \approx
\langle G^a_{4\alpha}(0)G^a_{4\alpha}(0)\rangle\equiv
-\langle {\cal E}^2\rangle \;,
\end{equation}
where ${\cal E}_i^a$ is the color electric field, we find
\begin{equation}
\langle G^a_4(x)_{np}G^a_4(x')_{np}\rangle\approx
-\frac{1}{12}\bm{x}\cdot\bm{x}'
\langle {\cal E}^2\rangle \;.
\end{equation}

Using the results above, the computation of $\langle \theta^2\rangle$ is long
but 
straightforward. Collecting all terms, and assuming a large volume, we find
\begin{eqnarray}
\langle \theta^2\rangle &\approx&
\frac{\mu^2N_f^2(N_c^2-4)}{(12\pi^2)^2N_c}
\left[
\frac{\pi\langle {\cal E}^2\rangle^2 g^2}{8(N_c^2-1)T}
\left(\frac{R^{7}}{m^2}+\frac{R^{5}}{m^{4}}  \right)
\right. \nonumber \\
&&\left. -\frac{3\langle {\cal E}^2\rangle g^4}{32\pi}\frac{R^{5}}{5m}
+ F(\beta m,\beta \Lambda_{\rm \overline{\mathrm{MS}}})Tg^{6} R^{3}
\right],
\end{eqnarray}
where $R$ is the radius of the system in spherical coordinates, 
$\Lambda_{\overline{\mathrm{MS}}}$ is the energy scale in the modified minimal
subtraction (MS) scheme, and 
$F$ is an integral that can be computed numerically.

The last term is purely perturbative, whereas the other ones come from the mixed
contribution. 
The purely nonperturbative contribution vanishes identically in the 
mean field approximation. Notice that, in the thermodynamic limit, the dominant 
contribution is the one proportional to $R^{7}\sim V^{7/3}$, so that $\langle \theta^2\rangle$ 
grows faster than quadratically with the volume, the other contributions being essentially 
irrelevant.

\section{Conclusions and outlook}

We presented a well-defined and systematic procedure for computing the fermionic 
determinant in QCD at finite temperature and chemical potential. In the framework 
defined by the power counting method described above, we calculated exactly the 
minimal corrections to the effective action. This procedure can be systematically 
extended to higher orders.

We also computed explicitly $\langle \theta^2\rangle$ in the high-temperature limit 
and analyzed its volume dependence. In this limit, the dominant contribution comes 
naturally from the zero Matsubara mode, so that dimensional reduction is a good 
approximation. Although we do not present results beyond the mean 
field approximation for the gluon condensate, we expect that in that case, a
new 
length scale associated with the nonuniformity of the condensate will compete
with 
the scale $R$ but should not modify the picture appreciably in the limit
considered 
in this work.

In this paper we proposed a new way of approaching the problem of determining
the phase of the complex functional determinant in QCD at small chemical potential and
relatively large temperatures, an important issue for lattice QCD thermodynamics.
Our framework clearly contains, and goes beyond, the dimensional reduction approximation,
being valid for temperatures closer to $T_{c}$. 
We believe that this analysis, complemented
by previous existing results, might shed some light onto the phase transition in
the region of high
temperature and low density presumably probed by relativistic heavy-ion
experiments, and perhaps
indicate a simple way to handle the sign problem for temperature values closer to the critical temperature.

\section*{Acknowledgment}

The authors acknowledge financial support from ANPCyT, CAPES, CLAF, CNPq, FAPERJ and
FUJB-UFRJ.


\end{document}